\documentclass[10pt,letterpaper]{article}
\usepackage[top=0.85in,left=2.75in,footskip=0.75in]{geometry}

\usepackage{amsmath,amssymb}

\usepackage{changepage}

\usepackage[utf8x]{inputenc}

\usepackage{textcomp,marvosym}

\usepackage{cite}

\usepackage{nameref}
\usepackage{hyperref}
\usepackage[right]{lineno}

\usepackage{microtype}
\DisableLigatures[f]{encoding = *, family = * }

\usepackage[table]{xcolor}

\usepackage{array}

\newcolumntype{+}{!{\vrule width 2pt}}

\newlength\savedwidth

\raggedright
\usepackage[aboveskip=1pt,labelfont=bf,labelsep=period,justification=raggedright,singlelinecheck=off]{caption}

\makeatletter
\renewcommand{\@biblabel}[1]{\quad#1.}
\makeatother

\usepackage{lastpage,fancyhdr,graphicx}
\usepackage{epstopdf}
\usepackage{amssymb}
\fancyheadoffset[L]{2.25in}
\fancyfootoffset[L]{2.25in}
\usepackage{float}

\begin{document}
\vspace*{0.2in}

\begin{flushleft}
{\Large
\textbf\newline{One step back, two steps forward: interference and learning in recurrent neural networks} 
}
\newline
\\
Chen Beer\textsuperscript{1,2},
Omri Barak\textsuperscript{2,3*},
\\
\bigskip
\textbf{1} Viterby Faculty of Electrical Engineering, Technion Israel Institute of Technology, Haifa, Israel
\\
\textbf{2} Network Biology Research Laboratories, Technion Israel Institute of Technology, Haifa, Israel
\\
\textbf{3} Rappaport Faculty of Medicine, Technion Israel Institute of Technology, Haifa, Israel
\\
\bigskip
* omri.barak@gmail.com

\end{flushleft}

\section*{Abstract}
Artificial neural networks, trained to perform cognitive tasks, have recently been used as models for neural recordings from animals performing these tasks. While some progress has been made in performing such comparisons, the evolution of network dynamics throughout learning remains unexplored. This is paralleled by an experimental focus on recording from trained animals, with few studies following neural activity throughout training. 

In this work, we address this gap in the realm of artificial networks by analyzing networks that are trained to perform memory and pattern generation tasks. The functional aspect of these tasks corresponds to dynamical objects in the fully trained network -- a line attractor or a set of limit cycles for the two respective tasks. We use these dynamical objects as anchors to study the effect of learning on their emergence. We find that the sequential nature of learning -- one trial at a time -- has major consequences for the learning trajectory and its final outcome. Specifically, we show that Least Mean Squares (LMS), a simple gradient descent suggested as a biologically plausible version of the FORCE algorithm, is constantly obstructed by forgetting, which is manifested as the destruction of dynamical objects from previous trials. The degree of interference is determined by the correlation between different trials. We show which specific ingredients of FORCE avoid this phenomenon. Overall, this difference results in convergence that is orders of magnitude slower for LMS.

Learning implies accumulating information across multiple trials to form the overall concept of the task. Our results show that interference between trials can greatly affect learning, in a learning rule dependent manner. These insights can help design experimental protocols that minimize such interference, and possibly infer underlying learning rules by observing behavior and neural activity throughout learning.


\section*{Introduction}
In recent years, trained recurrent neural networks were used as a model for several experimental tasks in neuroscience \cite{Mante2013,Miconi2017,Russo2018a,Remington2018,Wang2018,Sussillo2014,Barak2017}. In order to elucidate simple learning tasks, recurrent neural networks were designed \cite{Machens2005a,Miller2003,Ben-Yishai1995,Singh2006} and trained \cite{EminOrhan2017,Barak2013,Miconi2017,Song2017,Song2016} to memorize analog values or periodic patterns. In the resulting networks, memory was implemented by attractors in the neural state-space, in particular fixed points and limit cycles, respectively. The stability properties of these attractors were studied to understand where they arise from, how they can be improved \cite{itskov2011}, and how they link to experimental observations \cite{wimmer2014,Wang2018,Russo2018a}. While some progress was made on understanding network dynamics following training \cite{Mante2013,Remington2018,sussillo2015,Yang2019}, the formation of dynamics throughout learning remains a puzzle. 

Learning memory tasks entails two interacting dynamical processes. Within the learning of a single trial, the system is expected to create a stable attractor, representing the memorized target. On a longer timescale, learning the complete task requires integrating sequentially presented trials into a coherent behavior. Sequential learning from many trials was mostly studied in the context of learning multiple tasks; in this case, the danger of forgetting was stressed, and several suggestions were made on how to alleviate it \cite{kirkpatrick2017,Zenke2017,zeno2018}. However, forgetting within a single task -- from one trial to another -- was not explored, and neither the reasons for such cases.

We address these gaps -- formation of dynamics and forgetting within a single task -- in a simple setting. We train a recurrent neural network on two tasks: a simple memory task and a memory task combined with pattern generation. We use both FORCE learning \cite{sussillo2009a} and its simple, biologically plausible version -- LMS. By reverse engineering the networks, we are able to follow the formation of dynamical structures throughout training. We show that, depending on the learning rule used, the sequential nature of learning can cause major interference within a single task. By studying these dynamics, we reveal the underlying causes for training success and failure. Specifically, we show that LMS learning suffers heavily from interference between trials, causing the previous dynamical objects to vanish. We show that this phenomenon results from high correlations between trials, and expose the explicit components of the FORCE algorithm that prevent such destruction. We show that this fundamental difference between the two algorithms, contrary to previous beliefs \cite{sussillo2009a,Hoerzer2014}, results in orders of magnitude slower convergence of LMS compared to FORCE. Finally, we show how careful ordering of trials can decrease this interference and promote efficient sequential learning.

\section*{Results}

\begin{figure}[H]
  \centering
  \includegraphics[width = 1.0\textwidth]{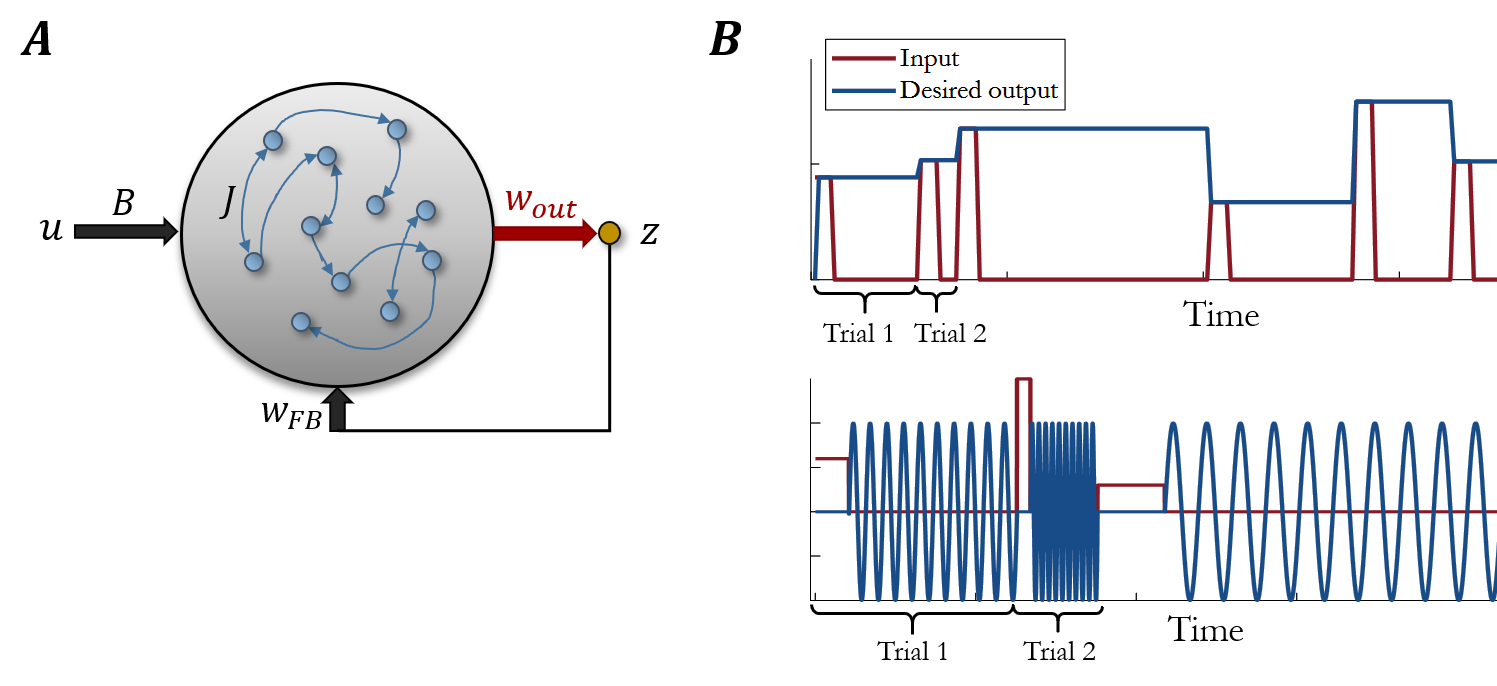}
  \caption{{\bf Network architecture and tasks illustration.}
  \textbf{(A)} Network architecture. Only the outputs weights are modified during training (see methods).
  \textbf{(B)} Top: Simple memory task. At each trial a stimulus with amplitude in the range $[1,5]$ is presented for 500 ms (red). The delay period between different stimuli is uniformly distributed between 0.5 and 6 seconds. The desired behavior is to output this value for the entire trial duration (blue). Bottom: Memory with pattern generation task. At each trial a DC stimulus is presented for 2 cycles (red). The desired behavior is to output sine and cosine waves matching to this frequency for 10 cycles (blue, only sine shown). Frequencies are in the range $[0.2,1.1]$.
  \label{fig:Architecture&Tasks}}
\end{figure}

We use a rate based recurrent neural network, defined by the equations (Fig \ref{fig:Architecture&Tasks}A):
\begin{equation}
    \tau \dot{x_i} = -x_i+\sum_{j=1}^{N}{J}_{ij}r_j+{B}_{i}u+w^{FB}_{i}z  ,
    \label{eq:xdot}
\end{equation}
where ${x}_{i}$ is the input to neuron $i=1,\ldots,N$, $J$ is a random connectivity matrix, $r_{i}=\phi(x_i)$ is the firing rate with $\phi(x)=tanh(x)$, and the external inputs $u$ are fed through weights $B_i$. The current state of the network is defined by $x\in \mathbb{R}^N$ and the output of the network $z= w_{out}^T r$ is fed back through weights $w^{FB}$. 

The network is trained on two memory tasks as follows: The first task is a simple memory task -- we present transient stimuli to the network, and train it to output their value for a long time after stimulus removal (Fig \ref{fig:Architecture&Tasks}B, top). The stimulus amplitudes are uniformly sampled from $[1,5]$, and are presented for 500 ms, followed by a delay period uniformly sampled from $[0.5,6]$ sec. 

The second task is a memory task combined with pattern generation -- the network is required to generate a sine wave after receiving a specific frequency value as an input for a limited period of time. The network receives the required frequency as a DC input for 2 cycles and produces the corresponding sine wave for 10 more cycles in the absence of input (Fig \ref{fig:Architecture&Tasks}B, bottom). For this task, we use a two dimensional output -- demanding both a sine and a cosine.

\begin{figure}[H]
  \centering
  \includegraphics[width = 0.75\textwidth]{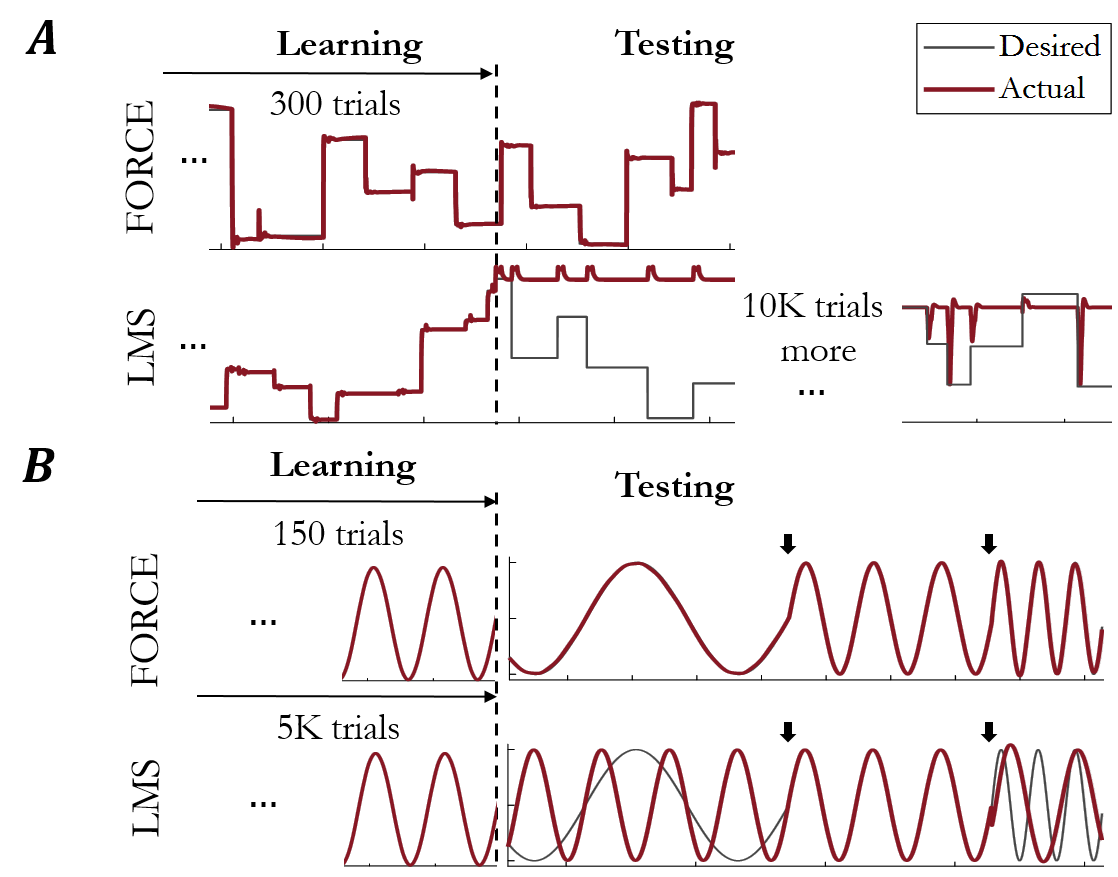}
  \caption{{\bf Learning the complete task.}
  Actual network output (red) using FORCE (top) and LMS (bottom) training. During learning, both algorithms lead to the desired output (grey). Testing, however, reveals that the LMS-trained network converges to the output value of the last training stimulus. The dashed line denotes the beginning of testing. The rightmost plot in \textbf{(A)} shows a test session following an additional $10^4$ training trials. \textbf{(B)} Arrows indicate target switch; testing only 3 cycles. 
  \label{fig:testing}}
\end{figure}

We trained networks on the simple memory task for 300 trials using either FORCE or LMS, modifying only output weights in both cases (Fig \ref{fig:Architecture&Tasks}A, Methods). Figure \ref{fig:testing}A shows that while FORCE training resulted in almost perfect performance, this was not the case for LMS. Closer inspection of the LMS output during testing indicates that the network converged to a fixed point corresponding to the last trained stimulus value. Since in \cite{sussillo2009a,Hoerzer2014} LMS was shown to require roughly 10 times more trials to converge than FORCE, we continued training for a further $10^4$ trials, but to no avail.

Figure \ref{fig:testing}B shows test results for the memory and pattern generation task -- FORCE training resulted in a network which was able to output all different frequencies perfectly, while LMS converged to the last training stimulus, as in the previous task.

\subsection*{Emergence of dynamics} 
To gain insight into the underlying reasons for this behavior, we reverse engineer the network to uncover its dynamics. 

In the simple memory task, in a successfully trained network, we expect to find an approximate line attractor -- a one dimensional manifold of fixed points. Each point along this line represents a memorized value. In the second task, for each learned frequency we expect to find a stable limit cycle. 

As we only train the output weights, the location of these attractors is pre-determined in the following manner. 
For a stimulus $s$ (value or frequency depending on the task) the desired output is $z_s(t) = A_s(t)$. We consider the open loop system (in which the target output $A_s$ is fed back to the network), and note that it converges to a unique stable trajectory $\bar{x}_s(t)$ which is given by

\begin{equation}
\bar{x}_s(t)=J\phi(\bar{x}_s(t))+w^{FB}A_s(t)  .
\label{eq:xbar}
\end{equation}

\begin{figure}[H]
  \centering
  \includegraphics[width = 1.0\textwidth]{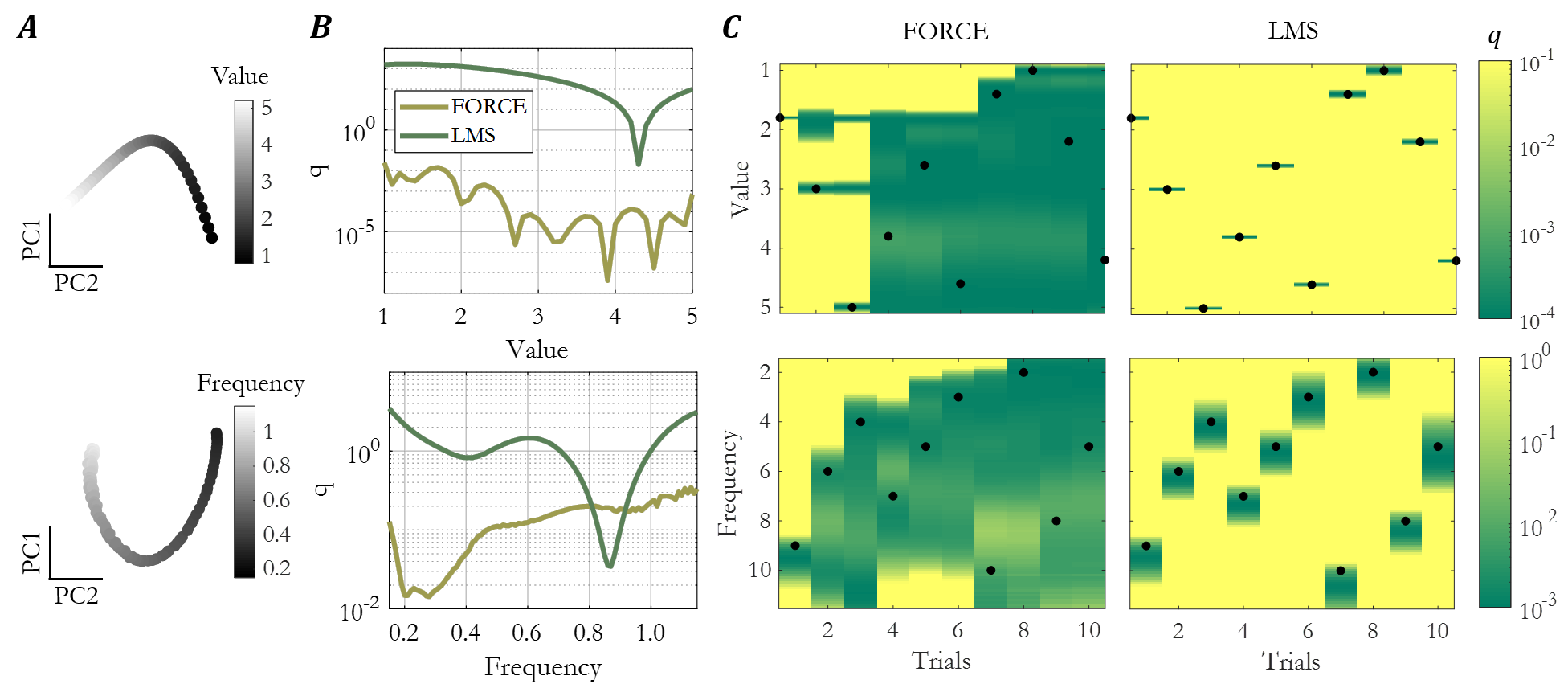}
  \caption{{\bf Emergence of an attractor.}
  \textbf{(A)} The locations $\bar{x}$ in PCA space for the entire output range $z\in[1,5]$ in the first task (top) and for the entire frequency range $\omega\in[0.2,1.1]$ at phase zero in the second task (bottom). \textbf{(B)} $q$-values along the expected attractor following training in either FORCE (yellow) or LMS (green). Top -- simple memory task: following 300 trials of training. Bottom -- memory and pattern generation task: following 150 trials of training. \textbf{(C)} $q$-values along the expected attractor during the first 10 trials for both FORCE and LMS. The black dots denote the specific stimulus at each trial. Simple memory task (top), memory and pattern generation task (bottom).
  \label{fig:PCA_q}}
\end{figure}

This happens for $A$s that are large enough to suppress chaos \cite{Rajan2010}, which is true for the values used in this work. Figure \ref{fig:PCA_q}A shows $\bar{x}_s(t)$ in PCA space for the entire output range $[1,5]$ in the simple memory task (top), and $\bar{x}_s(0)$ states (phase zero) for the entire output frequency range $[0.2,1.1]$ in the pattern generation task (bottom). We used zero phase because the centers of all limit cycles coincide (see appendix \nameref{Appendix_limitCycles})

Training the output weights can determine the existence and stability of these putative attractors in the closed loop system (as it introduces a low rank modification to the recurrent connectivity). Existence relies on a consistency between the actual and desired network outputs on the attractors. This consistency can be quantified by a scalar function \cite{sussillo2013}:

\begin{equation}
q_s=\frac{1}{2}\left| \int_0^T  w^{FB}(w'_{out}\phi(\bar{x}_s(t))-A_s(t)) dt\right|^2  ,
\label{eq:qs}
\end{equation}

which represents the difference between the closed loop and the open loop dynamics. Note that $q_s=0$ if and only if $\bar{x}_s$ is a fixed point (or a limit cycle for the second task), but stability is not guaranteed (see appendix \nameref{Appendix_stability}).

Figure \ref{fig:PCA_q}B shows the value of $q_s$ for different output values in the first task (top) and for different frequencies in the second (bottom). The resulting behavior of the FORCE and LMS trained networks is captured by these graphs. The $q$-values of FORCE are very low for the entire range, indicating an approximate line attractor in the first task and a set of stable limit cycles in the second. Conversely, LMS leads to a single attractor -- a single fixed point around $z=4.2$ in the first task and a single limit cycle around $\omega=0.93$ in the second, matching the behavior seen in Figure \ref{fig:testing}. When inspecting these curves during learning, the difference between the two algorithms is clearly demonstrated -- following each trial, a new attractor emerges in both FORCE and LMS but it remains to exist after the subsequent trial only in FORCE (Fig \ref{fig:PCA_q}C). 

\subsection*{Memory and interference}
What enables FORCE-trained networks to have many attractors, while LMS only leads to a single one? Our results suggest that both algorithms will lead to a fixed point after the first trial, but only FORCE will maintain this fixed point after the second trial. 
The FORCE algorithm, through its use of recursive least squares, uses second order information \cite{Vaits,Gruber1997}. The Hessian, which is the second order derivative of the loss, is given by:
$\frac{\partial^2}{\partial w^2}\frac{1}{2}(w^T r-f)^2 = r r^T$, 
and is stored in $P$. $P$ is a running estimate for the inverse correlation matrix of the network rates $r$ plus a regularization term:
\begin{equation}
P(t) = \left(\int_{0}^{t} r(\tau)r^T(\tau)d\tau+\alpha I\right)^{-1} .
\label{eq:P}
\end{equation}
The eigenvectors of $P$ are thus approximately the principal components of the activity, but since $P$ retains the inverse of the activity history, the eigenvalues of these principle components will be very small. The update of synaptic weights in FORCE is proportional to $P$ (Eq \ref{eq:RLS_LR}), and thus protects the directions most visited (e.g. fixed points and points along the limit cycles) from further change.

Now that we understand the information stored in $P$, we can examine what happens during the second trial. We denote the targets of the first two trials $A_1$ and $A_2$ with the corresponding putative attractors $\bar{r}_1$ and $\bar{r}_2$. 
In the simple memory task (Fig \ref{fig:PCA_q}C, top), training with FORCE results in two stable fixed points, indicating that $r_1$ was preserved. Training with LMS causes the network to forget the first value, and as a result, only one fixed point corresponding to the last value $A_2$ remains. Correspondingly, in the pattern generation task (Fig \ref{fig:PCA_q}C, bottom), training with FORCE results in two stable limit cycles, while in LMS after the second trial there is a single limit cycle matching the last stimulus frequency. 




To prove that the second order information is responsible for avoiding interference between trials, we reset the $P$ matrix after each trial to be $(\alpha I)^{-1}$. The results are virtually identical to those of LMS (see appendix \nameref{Appendix_P_reset}), confirming our hypothesis. Because the second order information consists of correlations between different attractors, we inspect the effect of these correlations more carefully.

In our setting, the rate vectors corresponding to different attractors (Fig \ref{fig:PCA_q}A) are highly correlated. Therefore, when learning these rate vectors sequentially, first order algorithms like LMS suffer from constant interference, leading to what seems to be a complete forgetting of the past. Inspecting this phenomenon thoroughly uncovers the slow underlying processes.

\begin{figure}[H]
  \centering
  \includegraphics[width = 1.0\textwidth]{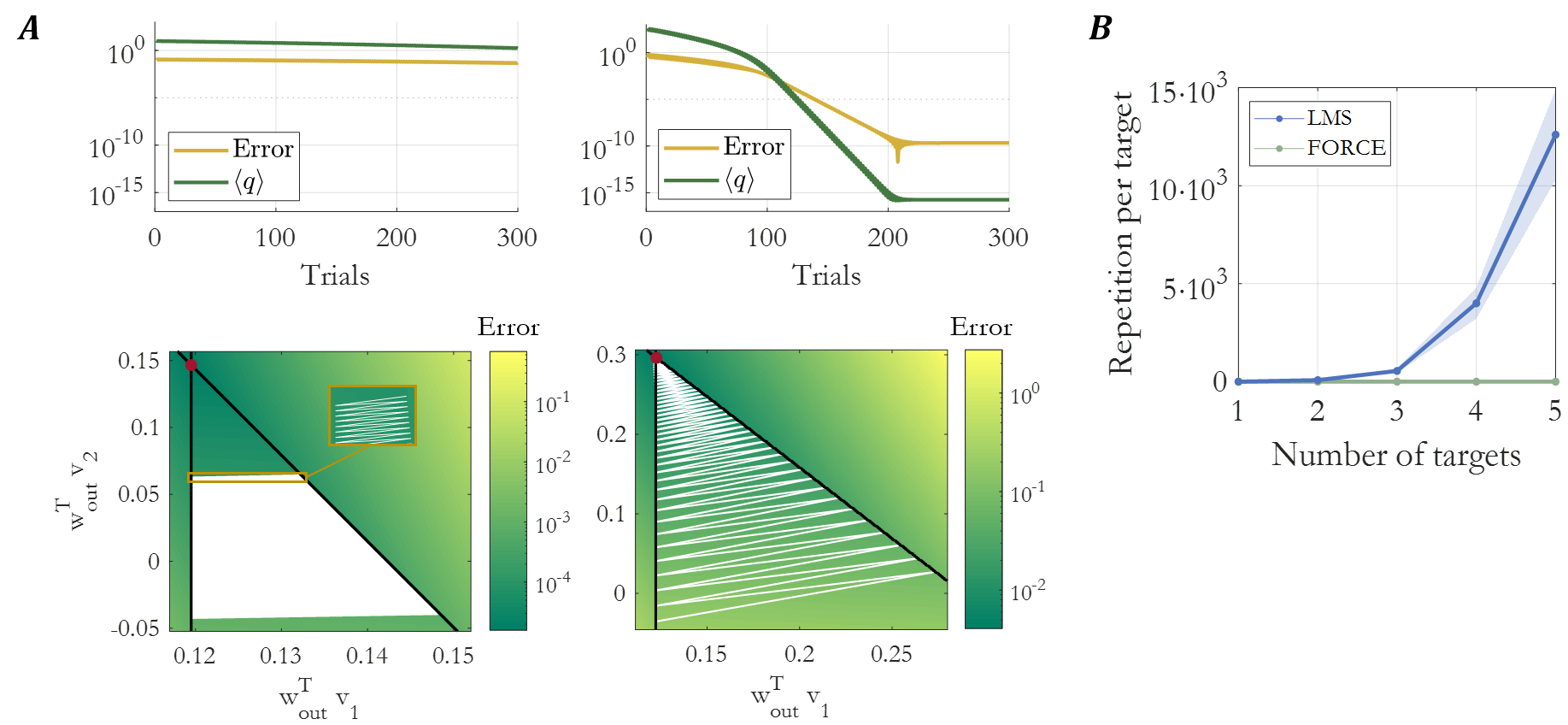}
  \caption{{\bf Interference in LMS.}
  \textbf{(A)} Simple memory task for only two values $A_1, A_2$ repeatedly presented while learning using LMS. Top: average $q$-values measured at the corresponding fixed points $\bar{r}_1, \bar{r}_2$ (green) and the validation error during learning (yellow); left: $A_1=2, A_2=2.5$, right: $A_1=2, A_2=5$. Bottom: the projections of $w_{out}$ onto $v_1=\bar{r}_1$ and $v_2=\bar{r}_2-\frac{\langle \bar{r}_1,\bar{r}_2 \rangle}{\langle \bar{r}_1,\bar{r}_1 \rangle}\bar{r}_1$ (white); background colors represent the average error, the red dot denotes the optimal solution, and the black lines are the local minima for each target value. The inset on the left highlights the destructive interference at each step in the case where $\bar{r}_1$ and $\bar{r}_2$ are more correlated.
  \textbf{(B)} Number of trials per target required to achieve 1\% error as a function of the number of targets for LMS (blue) and FORCE (green). Target values are sampled from $[1, 5]$; training continues as long as the error is higher than 1\%. The solid line denotes the mean of 5 simulations. Light blue shade denotes the standard deviation across simulations. 
  \label{fig:LMS_interference}}
\end{figure}

We consider the example of memorizing only two target values, this time for two cases that differ in the correlation between $\bar{r}_1$ and $\bar{r}_2$. When the correlation is high, viewing the validation error throughout learning might imply that no learning is taking place (Fig \ref{fig:LMS_interference}A, left, yellow line). On the other hand, observing the projections of $w_{out}$ onto the plane spanned by the corresponding fixed points $\bar{r}_1, \bar{r}_2$ reveals the slow underlying process. The black lines in Figure \ref{fig:LMS_interference}A show zero error for either target, and their intersection (red dot) is the optimal solution. The training trajectory (white) reveals that learning is constantly interfered (Fig \ref{fig:LMS_interference}A, bottom), but the independent components of $\bar{r}_1$ and $\bar{r}_2$ survive. The network converges eventually when training one target no longer affects the readout from the other, and the following equalities hold: $w_{out}^T \bar{r}_1 = A_1 $, $w_{out}^T \bar{r}_2 = A_2 $. When $\bar{r}_1$ and $\bar{r}_2$ are less correlated, the amount of interference decreases, leading to much faster convergence (Fig \ref{fig:LMS_interference}A, right). 

The complete task of memorizing a wide range of values requires the formation of a line attractor -- a continuous set of highly correlated fixed points. Figure \ref{fig:LMS_interference}B shows the number of trials required for LMS and FORCE as the number of target values increases. The detrimental effects of interference rapidly add up and lead to diverging convergence times.\newline

The interference shown in figure \ref{fig:LMS_interference}A -- constant alternation between local minima -- might appear as an outcome of a very high learning rate. This is also consistent with the observation that an attractor is formed following a single trial (Fig \ref{fig:PCA_q}C). We therefore speculated that decreasing the learning rate would lead to a learning trajectory which converges directly to the optimal solution with no oscillations. One can consider this case as equivalent to learning several trials in parallel (e.g. batch learning). Unfortunately, in this slow learning regime, control issues arise -- learning is too slow to keep the network on the open-loop manifold within a trial. This results in a gradient which is not in the correct direction (see appendix \nameref{Appendix_control}).

\subsection*{Minimizing interference}
In both the training of animals (shaping, \cite{Skinner1958}) and of artificial networks (curriculum learning, \cite{Bengio2009}), it has been observed that the order of training can have an effect on the final performance. Our results indicate an interplay between the correlation of sequentially learned fixed points and the process of learning. We hypothesize that this feature can be used to design optimal training sequences.

\begin{figure}[H]
  \centering
  \includegraphics[width = 1.0\textwidth]{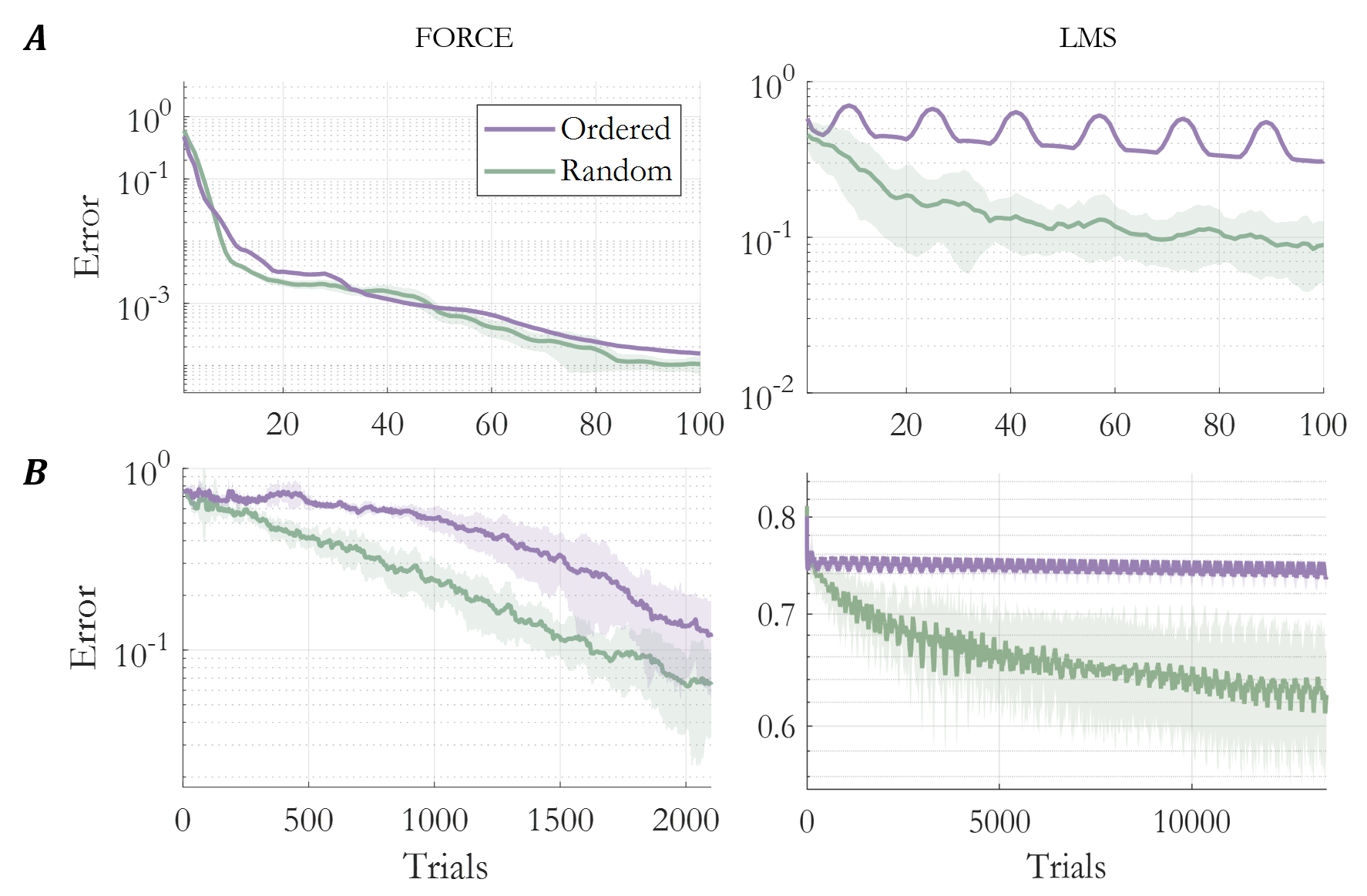}
  \caption{{\bf Order of trials affects performance.}
  Average error along the learning process for two learning sequences -- random (green) and ordered (purple) for both FORCE and LMS. The solid lines denote the mean of 10 simulations; light shades denote the standard deviation across simulations.\textbf{(A)} Simple memory task using 9 stimulus values from $[1,5]$ \textbf{(B)} Memory and pattern generation task, using 10 values from $[0.2,1.1]$ (See methods).
  \label{fig:history}}
\end{figure}

In the two-target training case above, the correlation between consecutive stimuli was given by the choice of targets. When learning a full line attractor or a set of limit cycles, however, the precise sequence matters. As a proof of concept, we compare learning when the target values and frequencies are presented in an ordered fashion (Fig \ref{fig:history}, purple) or in a random ordering (green). The ordered sequence has very high correlations between neighboring targets. The random sequence is expected to be more balanced, with lower correlation values. For each ordering, we can calculate the sum of correlation coefficients along the sequence. Indeed, considering the maximal and minimal sums attainable, the former is obtained by the ordered sequence, whereas random sequences result in intermediate values. Based on our previous results (Fig \ref{fig:LMS_interference}A), we expect these sums to affect the level of interference during learning. When using LMS, training with a random sequence proceeds much faster compared to the ordered sequence (though the error still remains high after many trials). In this case, each new attractor is sufficiently different from the previous one, and therefore the interference during learning is smaller. Moreover, when comparing the error decay of a random sequence training to the one obtained by the minimal sum sequence -- the results are very similar (see methods).
As demonstrated in previous sections, FORCE is able to protect previously visited attractors, and therefore, the difference between random and ordered training sequences is much smaller than in LMS learning.

\section*{Discussion}


In this work we followed the formation of dynamical objects through the online training of a recurrent neural network. To enable a feasible analysis we designed two simple memory tasks, each one requiring a different dynamical object in the resulting dynamics. We showed that FORCE, an RLS based algorithm, avoids forgetting by implicitly avoiding updates to previously visited directions. In contrast, we showed that LMS learning suffers from constant interference, leading to recurring elimination of what was learned in earlier stages. By dissecting the contributions of the various parts of a trial to the implicitly acquired memory, we could understand how the formation and maintenance properties of these dynamical objects lead to learning. 

We focused on the FORCE algorithm due to its simplicity and usage within neuroscience inspired tasks \cite{Laje2013a,Barak2013}. This algorithm, however, is not biologically plausible. Both LMS and a reward based rule \cite{Hoerzer2014} were suggested as slower but effective plausible versions of FORCE. Our results indicate that this might not be the case in more complex tasks (for the reward based rule see appendix \nameref{Appendix_Hoerzer}). We demonstrate the qualitative differences between FORCE and LMS learning mechanisms, leading to orders of magnitude difference in convergence rate, which rapidly increases with task complexity.

Aiming to understand the learning process of cognitive tasks, we addressed the case of online learning, in which learning takes place continuously while the trials are presented sequentially. This differs from the typical machine learning setting of batch learning, in which several stimuli are presented in parallel. In our setting, online learning is manifested by the formation of a new dynamical object after each trial either in FORCE or LMS. When using very small learning rates, such dynamics are not formed after a single trial, and therefore online learning could be equivalent to presenting multiple trials at the same time. In this case, however, control problems arise as learning is too slow to force the network dynamics in the correct direction. 

Although we stress the sequential nature of learning, the fact that this is a single task leads to important distinctions from the sequential learning of uncorrelated tasks. In the simple memory task, the rate vectors corresponding to different points along the line attractor are highly correlated, and the same holds for the pattern generation task. Our results show that these correlations are the main cause for the extensive interference during LMS learning, seemingly leading to catastrophic forgetting within a single task.

Shaping, or choosing an optimal sequence of training examples, is a topic of importance in both animal training and machine learning research \cite{Skinner1958,Bengio2009}. Here, we show that correlations between sequential trials could indicate difficulty, and suggest to decrease interference within trials and accelerate learning by minimizing these correlations. In the future, it might be possible to probe the system to understand the correlation properties of neural representations of stimuli and use this information to guide the ordering of the following stimuli to reduce interference during training. Our results indicate no significant difference between random and optimal ordering, therefore, these online improvements might be simple to achieve. Such an online choice of the next training target is similar in spirit to adaptive choice of stimuli that was suggested as an efficient method to estimate response properties of neurons \cite{Pillow}. 

Understanding the dynamics of trained recurrent neural networks is an important and difficult challenge \cite{Sussillo2014,Gao2015,Barak2017}. Understanding the formation of these dynamics is an even more formidable task. Here, we take a first step towards this goal and show that it sheds light on the nature of learning rules, and can provide clues on how to improve learning.

\section*{Methods}

\textbf{Learning rules. }
We inspect the learning process using two different training methods:
FORCE algorithm \cite{sussillo2009a}, based on recursive least squares (RLS), with the following update rule:
\begin{eqnarray}
w(t) &=& w(t-\Delta t)-e^{-}(t)P(t)r(t)  ,\\
e^{-}(t) &=& w^{T}(t-\Delta t)r(t)-f(t)  ,
\label{eq:RLS_LR}
\end{eqnarray}
where $P(t)$ is a running estimate of the inverse correlation matrix of the network rates $r$ plus a regularization term, and $f(t)$ is the target output.

Although it is widely used, FORCE is not biologically plausible since the modification of a given synapse depends on information from the entire neural population. These locality considerations led the authors of \cite{sussillo2009a} to suggest a local learning rule:
Least mean squares (LMS), in which the modification rule for the output weights is 
\begin{equation}
w(t) = w(t-\Delta t)-\eta(t)e^{-}(t)r(t)  ,
\label{eq:LMS_LR}
\end{equation}
where $e^{-}(t)$ is defined as in FORCE, and $\eta(t)$ is a time-varying learning rate: $d\eta/dt=\eta(-\eta+|e^{-}|^\gamma)$.
A further step towards biological realism was made by the introduction of a reward based learning rule~\cite{Hoerzer2014}. This rule was described as leading to convergence on the same set of tasks as FORCE, albeit with a convergence rate that is slower than FORCE and similar to LMS.

We consider the case where only the output weights are modified during training (Fig \ref{fig:Architecture&Tasks}A, red), which facilitates our analysis. Because of the feedback connection, training still modifies network dynamics, and thus allows successful training of the task. Specifically, in the simple memory task the network outputs a single value, therefore training generates a rank one modification to the connectivity. In the pattern generation task the network is trained for both sine and cosine, thereby generating a rank two modification to connectivity. We also verified that the main effects are qualitatively similar when training the internal connections as well. \newline

\textbf{Simulation parameters. }
In all simulations $J$ elements were independently sampled from $\mathcal{N}(0,g^2/N)$ with $N=500$ and $g=1.2$. The input weights $B$ and the feedback weights $w^{FB}$ were drawn from a uniform distribution between $-1$ and $1$. $w_{out}$ were initially set to zero. $\tau = 100$ ms.
In FORCE: $\alpha = 10$. For LMS, we used $\gamma$ values of $1,2,5$, and found that the results were qualitatively similar to a constant $\eta$. Therefore, to reduce the number of parameters, we used constant $\eta$ for all figures.  \newline

\textbf{Figure \ref{fig:LMS_interference}B. } 
Networks were trained using both LMS and FORCE to perform the simple memory task for 1,2,3,4 and 5 target values. In order to disambiguate the pure formation and maintenance of fixed points from input pulse effects and transitions between fixed points, we use a reduced setup of the task. Instead of using external inputs to shift between the various output levels, we use the fact that the internal state $\bar{x}_s$ is precisely defined by output values $A_s$. We thus use these states as initial conditions for each trial, and eliminate external input.

Training stopped when reaching 1\% error difference between $w_{out}^T\bar{r}_s$ and the desired output $A_s$. FORCE training converges after a single trial for each value, while the number of trials required for LMS grows exponentially with the number of targets. Results shown for LMS were obtained using the fastest learning rate that converged to the desired solution: For a single target training $\eta=10^{-2}$ and for 2,3,4 and 5 targets $\eta \in [10^{-5}, 10^{-3}]$. Target values were [1], [1,2], [1,2,3], [1,2,3,4], [1,2,3,4,5].\newline

\textbf{Figure \ref{fig:history}. }
Networks were trained using both LMS and FORCE to perform the simple memory task for 9 target values in the range $[1,5]$ (A) and the pattern generation task for 10 frequencies in the range $[0.2, 1.1]$ (B). For each task, two training sequences were assessed -- a random sequence and an ordered sequence (1 to 5 and back to 1 repeatedly for the first task (A) and $\omega=0.2$ to 1.1 and back to 0.2 for the second (B)). In both cases, each target was presented at least 10 times.  

In addition to these two sequences, we tested the minimal-sum sequence: For a given set of target values or frequencies, we can find the corresponding attractors and calculate the correlation coefficients between them. For a single training round, in which each target is presented exactly once, we can find the sequence which results in the minimal correlation sum from one trial to the next. When presenting this sequence 10 times repeatedly, the validation error throughout training is very similar to the one obtained using a random sequence.

\section*{Appendix}

\paragraph*{A.1}
\label{Appendix_limitCycles}
{\bf A set of limit cycles. } 
\begin{figure}[H]
  \centering
  \includegraphics[width = 0.5\textwidth]{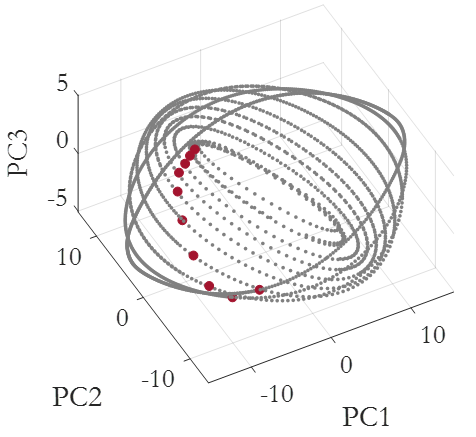}
  \caption{{\bf A set of limit cycles.}
  The locations $\bar{x}$ in PCA space for the entire frequency range $\omega\in[0.2,1.1]$ (grey), each limit cycle corresponds to a specific frequency. The red dots denote the phase zero state for each limit cycle.
  \label{fig:A1}}
\end{figure}

\begin{figure}[H]
  \centering
  \includegraphics[width = 1.0\textwidth]{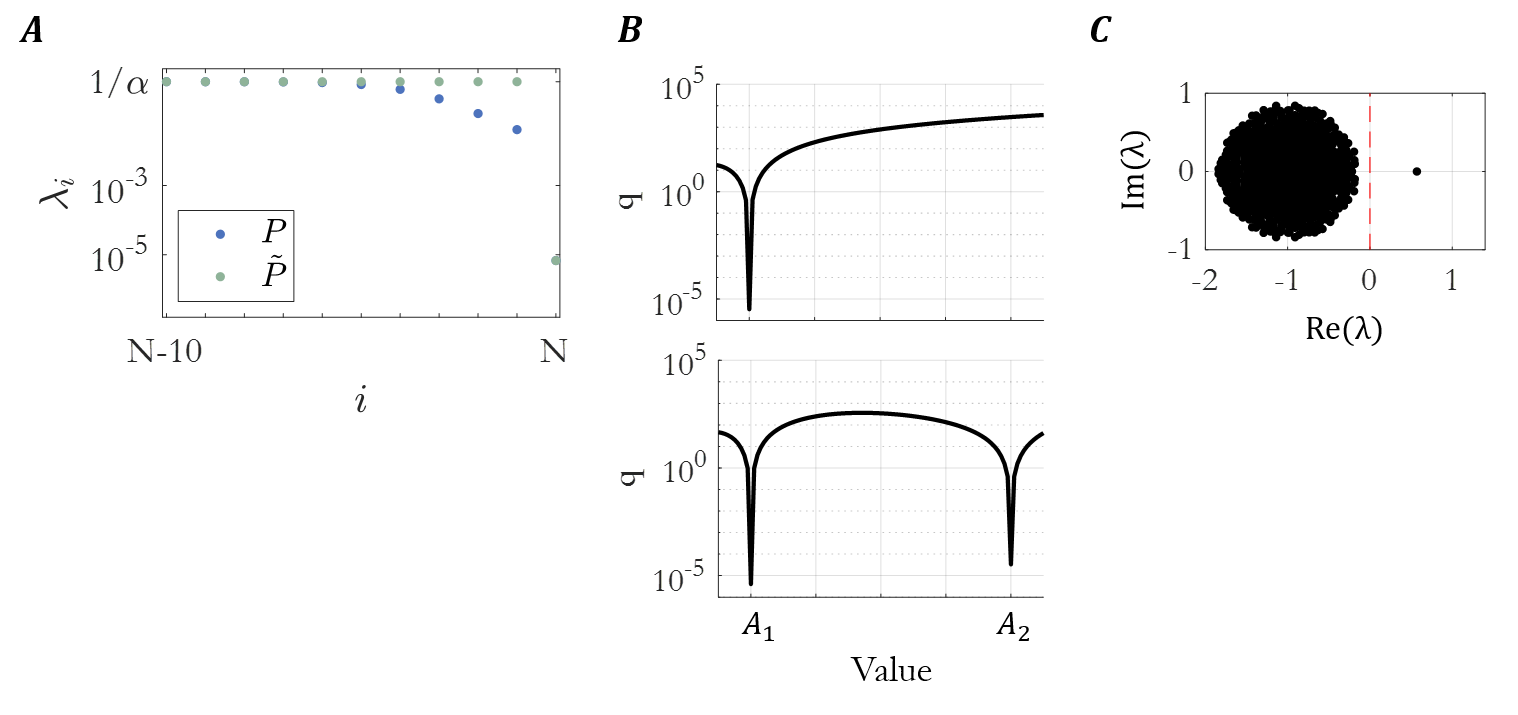}
  \caption{{\bf Partial inverse correlation matrix.} 
  \textbf{(A)} The last 10 eigenvalues of $P$ and $\tilde{P}$ following FORCE training of the first trial. The remaining $N-10$ equal $1/\alpha$ in both cases. \textbf{(B)} Top: $q$-values after the first trial ($A_1$), bottom: $q$-values after the second trial ($A_2$). \textbf{(C)} The spectrum of the network linearized around $\bar{r}_1$ after the second trial.
  \label{fig:A2}}
\end{figure}

\paragraph*{A.2}
\label{Appendix_stability}
{\bf Existence and stability. } 
In our simple memory task, each trial begins with a short transient until the network converges into the fixed point corresponding to the current target value. Assuming that the time spent at the fixed point $\bar{r}$ is significantly longer than the transient period, we can write the following approximation for $P$ following the first trial:
\begin{equation}
\tilde{P} = (T{\bar{r}\bar{r}^T+\alpha I})^{-1} ,
\label{eq:P_tilde}
\end{equation}
\noindent where $T$ is trial duration. The eigenvalues of $P$ and $\tilde{P}$ after a single trial are shown in Fig~\ref{fig:A2}A. The lowest eigenvalue in both matrices is $(\alpha + T\bar{r}^T\bar{r})^{-1}$, which corresponds to the eigenvector $\bar{r}$. The transient history is reflected in the eigenvalues of the nominal $P$, and not in $\tilde{P}$.
 
Performing training of two trials ($A_1=1, A_2=5$) using $\tilde{P}$, which contains information on the fixed point $\bar{r}_1$, but not the transients leading to it, results in an intermediate behavior -- the resulting network contains two fixed points, but only the one corresponding to the second trial is stable.

\begin{figure}[H]
  \centering
  \includegraphics[width = 1.0\textwidth]{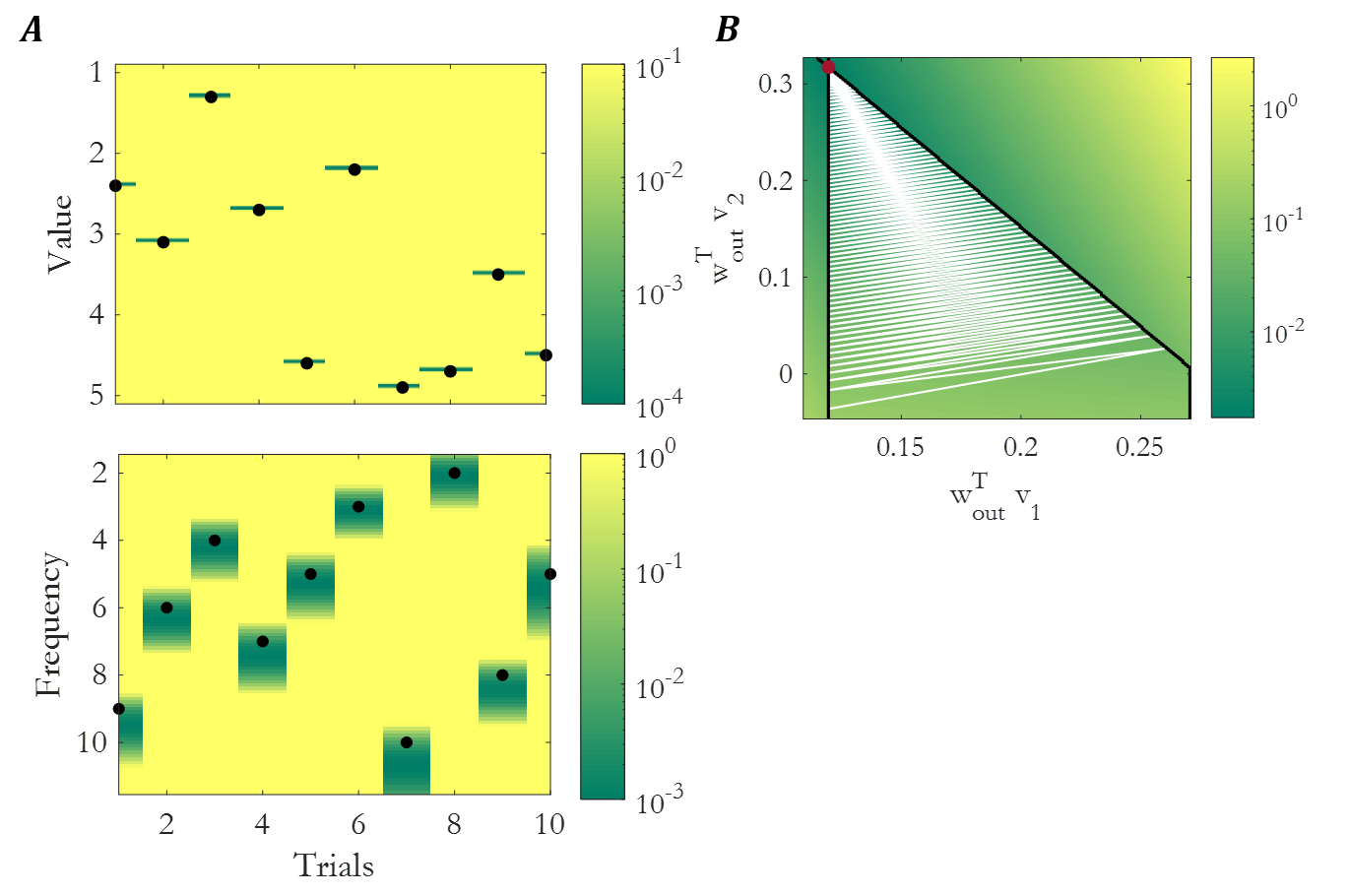}
  \caption{{\bf Deleting second order information between trials.}
  \textbf{(A)} $q$-values along the expected attractor during the first 10 trials. The black dots denote the specific stimulus at each trial. Top: Simple memory task, bottom: memory and pattern generation task.
  \textbf{(B)} The projections of $w_{out}$ onto $v_1=\bar{r}_1$ and $v_2=\bar{r}_2-\frac{\langle \bar{r}_1,\bar{r}_2 \rangle}{\langle \bar{r}_1,\bar{r}_1 \rangle}\bar{r}_1$ (white); background colors represent the average error, the red dot denotes the optimal solution, and the black lines are the local minima for each target value.
  \label{fig:A3}}
\end{figure}

\paragraph*{A.3}
\label{Appendix_P_reset}
{\bf Resetting the inverse correlation matrix. } 
To check whether the second order information is responsible for avoiding interference between trials, we reset the $P$ matrix to be $(\alpha I)^{-1}$ after each trial. The results shown in Figure \ref{fig:A3} are qualitatively identical to those of LMS learning, confirming that second order information allows learning with no interference between trials. 

\begin{figure}[H]
  \centering
  \includegraphics[width = 0.65\textwidth]{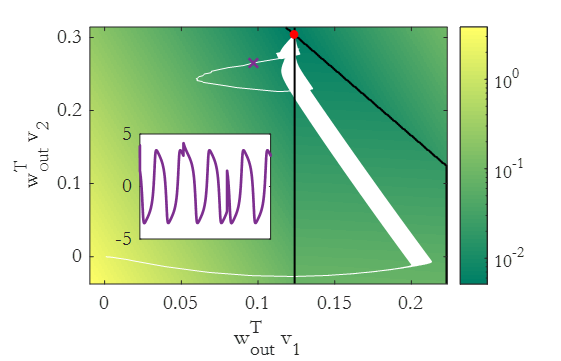}
  \caption{{\bf Control issues emerge.}
   The projections of $w_{out}$ onto $v_1=\bar{r}_1$ and $v_2=\bar{r}_2-\frac{\langle \bar{r}_1,\bar{r}_2 \rangle}{\langle \bar{r}_1,\bar{r}_1 \rangle}\bar{r}_1$ (white); background colors represent the average error, the red dot denotes the optimal solution, and the black lines are the local minima for each target value. Inset: the output of the network during learning around the position marked in a purple marker. Note that gradient descent operating around the purple marker guides learning in a direction which causes the error to increase.
  \label{fig:A4}}
\end{figure}

\paragraph*{A.4}
\label{Appendix_control}
{\bf Slow learning and control. } 
We explored learning using smaller learning rates ($\eta\in[10^{-8}, 10^{-6}]$). In this regime, each trial does not result in a new attractor. Initially, learning progresses along the boundary $w'_{out}\bar{r}_1=w'_{out}\bar{r}_2$ towards the optimal solution. However, in later stages, control issues arise as the network dynamics departs from the open-loop manifold vicinity. This effect exacerbates as the learning rate decreases. Figure \ref{fig:A4} illustrates the case of two targets learning in this slow learning regime ($\eta=10^{-6}, A_1=2, A_2=5$). The inset shows the output of the network during training in these extraneous areas, proving that indeed learning is too slow to force the dynamics to meet the desired behavior.

\paragraph*{A.5}
\label{Appendix_Hoerzer}
{\bf Reward based Hebbian learning. }
We trained networks on the two memory tasks described in the results section using a reward based learning rule~\cite{Hoerzer2014}. We used the same parameter settings detailed in that paper.

As demonstrated in the figure below, the network dynamics in both tasks is similar to the LMS-trained network.

\begin{figure}[H]
  \centering
  \includegraphics[width = 0.7\textwidth]{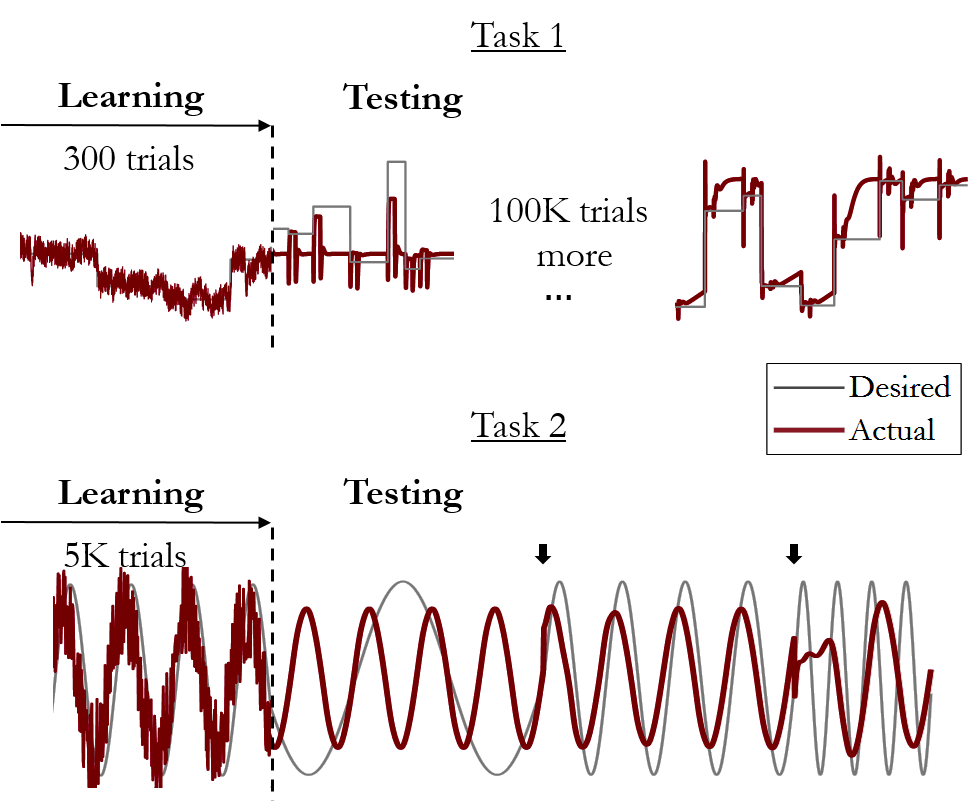}
  \caption{{\bf Reward based Hebbian learning.} 
  The actual network output (red) and the desired output (grey). Testing reveals that the network converges to the output value of the last training stimulus, similar to LMS. The dashed line denotes the beginning of testing. The rightmost plot in task 1 shows a test session following an additional $10^5$ training trials. Arrows in task 2 indicate target switch.
  \label{fig:A5}}
\end{figure}

\section*{Acknowledgments}
We thank Friedrich Schuessler and Lee Susman for comments on the manuscript.

\bibliographystyle{plos2015} 
\bibliography{mybib}

\end{document}